\documentclass[10pt,letterpaper]{article}
\usepackage{opex3}
\usepackage{color}
\usepackage{cite}
\usepackage{amsmath}
\usepackage{threeparttable}

\begin{document}

\title{Generalized circuit model for coupled plasmonic systems}

\author{Felix Benz,$^1$ Bart de Nijs,$^1$ Christos Tserkezis,$^2$ Rohit Chikkaraddy,$^1$ Daniel O. Sigle,$^1$ Laurynas Pukenas,$^3$ Stephen D. Evans,$^3$ Javier Aizpurua,$^2$ and Jeremy J. Baumberg$^{1,*}$}

\address{$^1$NanoPhotonics Centre, Cavendish Laboratory, JJ Thompson Ave, University of Cambridge, Cambridge, CB3 0HE, UK\\
$^2$Donostia International Physics Center (DIPC) and Centro de F\'{i}sica de Materiales, Centro Mixto CSIC-UPV/EHU, Paseo Manuel Lardizabal 4, 20018 Donostia/San Sebasti\'{a}n, Spain\\
$^3$Molecular and Nanoscale Physics, School of Physics and Astronomy, University of Leeds, Leeds, LS2 9JT, UK}

\email{$^*$jjb12@cam.ac.uk} 



\begin{abstract}
We develop an analytic circuit model for coupled plasmonic dimers separated by small gaps that provides a complete account of the optical resonance wavelength. Using a suitable equivalent circuit, it shows how partially conducting links can be treated and provides quantitative agreement with both experiment and full electromagnetic simulations. The model highlights how in the conducting regime, the kinetic inductance of the linkers set the spectral blue-shifts of the coupled plasmon.
\end{abstract}


\section{Introduction}

Plasmonic nanostructures are at the heart of many research fields ranging from biological applications \cite{Homola08,Taylor14} to quantum information processing \cite{Martino14,Martino12}. In recent years significant progress has been made in both the chemical preparation of noble-metal nanostructures \cite{Itoh04} and in simulation tools to model them \cite{Abajo97,Esteban12,Salandrino06,Pelton08,Taflove05}. However, few attempts have been made to develop a comprehensive analytical model \cite{Dhawan09,Trivedi14} that could facilitate direct validation of experimental results and yield parameters that can guide the implementation of full electromagnetic simulations. Most analytical approaches approximate Maxwell’s equations and are limited to specific geometries, like large gap separations \cite{Mazets00}. A promising alternative is to treat any nanoplasmonic system as a high frequency circuit composed of capacitors, inductors, and resistors \cite{Engheta05,Engheta07,Liu13,Shi14}. Such a circuit model is extremely versatile as generalized equations for the resonance condition can be obtained, and their dependence on the geometry of the nanostructure understood explicitly. Here we use this circuit approach to develop a simple analytical expression for two coupled plasmonic systems in close proximity. Surprisingly this has been absent until now, despite its extreme utility, because a number of subtle effects need to be considered. Our model is capable of describing both insulating and conductive gaps and applicable to arbitrary coupled plasmonic systems. We stress that such models are extremely useful to guide experiments, as electromagnetic simulation tools do not give direct insights (like those shown here) or scaling rules on the details of the nano-geometry. In the ultrasmall gap regime used here (and of strong interest for experiments), we show these models are accurate enough to be highly effective.

In this work we first review the circuit model for single nanoparticles, and show how to effectively account for nanoparticle size. We then show the full model for coupled plasmonics components when the gap between them is insulating, and when the gap has a finite conductance. We then show the origin of the scaling in the resonance coupled wavelength arising from the inductance of the gap region, producing a simple estimate for the blueshifts observed.

\section{Single particles}
\begin{figure}[htbp]
	\centering
	\includegraphics[width=7cm]{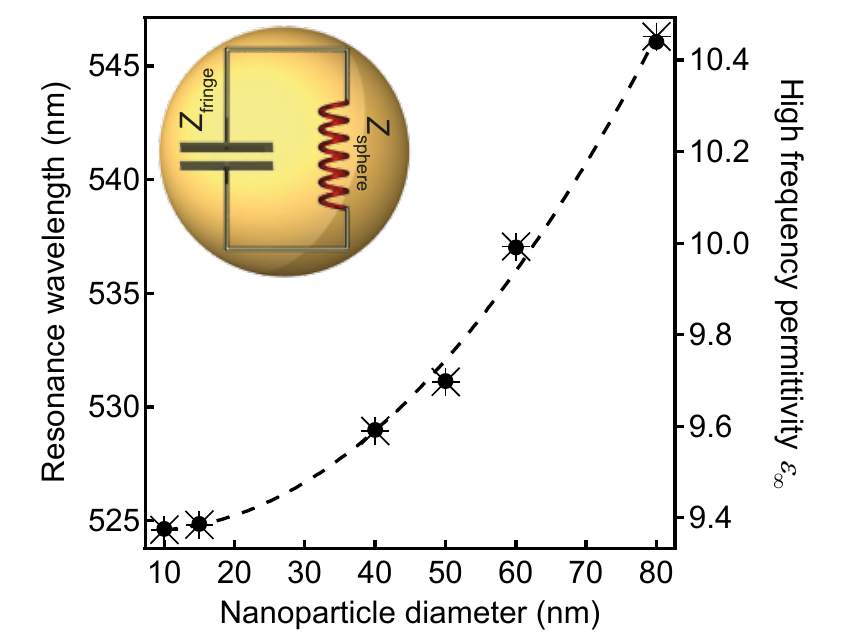}
	\caption{Resonance wavelength for different nanoparticle sizes and resulting values for $\epsilon_\infty$, determined from extinction spectroscopy in water. Inset: Single nanoparticle equivalent LC-circuit. Source data available in Dataset 1 (Ref.~\cite{OpenData}).}
	\label{Fig:1}
\end{figure}

We start from the description of a single spherical nanoparticle, which is well characterized by an LC circuit that has been shown to be the exact solution to Maxwell's equations. Engetha \textit{et al.}\cite{Engheta05} have determined the impedance contributions of the fringing field $Z_{\mathrm{fringe}}^{-1} = -\mathrm{i} \omega C_s = -\mathrm{i} 2 \pi \omega R \epsilon_0$ and the sphere $Z_{\mathrm{sphere}}^{-1} = -\mathrm{i} \pi \omega R \epsilon \epsilon_0$ with nanoparticle radius $R$, frequency $\omega$, gold permittivity $\epsilon$, vacuum permittivity $\epsilon_0$, and the capacitance due to the fringing field $C_s$. The resonance wavelength can be determined by $\Im (1/Z_{tot} )=0$, which yields the well-known resonance condition in the electrostatic case for a plasmonic nanoparticle $\epsilon = -2$. Here, throughout we use a Drude model for the dielectric function of gold:

\begin{equation}
\label{Eq:1}
\epsilon = \epsilon_{\infty}-\left(\frac{\omega_p}{\omega}\right)^2 = \epsilon_{\infty} - \left(\frac{\lambda}{\lambda_p}\right)^2.
\end{equation}
\noindent
This yields the single particle resonance frequency, $\lambda_s = \lambda_p (\epsilon_\infty + 2)$. The plasma frequency and wavelength ($\omega_p$ and $\lambda_p$ respectively), and the background high frequency permittivity ($\epsilon_\infty$) are material parameters which should not depend on nanoparticle size. However, electrodynamical effects, beyond the electrostatic approximation, lead to a weak particle size dependence of the resonance wavelength. There are several ways of correcting the resonance condition due to retardation \cite{Myroshnychenko08,Sarychev00}, however, a relatively simple phenomenological way to account for this is to assume that $\epsilon_\infty$ depends slightly on the nanoparticle size. The resonance wavelength of different gold nanoparticle sizes in water (permittivity $\epsilon_m = 1.33$) from extinction spectroscopy (Fig.~\ref{Fig:1}) provides estimated values for $\epsilon_\infty (R)$ using $\epsilon = -2 \epsilon_m$ together with Eq.~(\ref{Eq:1}), as we show in Dataset 1 (Ref.~\cite{OpenData}).

\section{Coupled plasmonic systems}
To describe purely capacitive-coupled resonances a capacitor (with capacitance $C_g$) can be placed between two equivalent single particle circuits mimicking the capacitive coupling, which reproduces the strongly red-shifted resonance \cite{Jain07, Sonnichsen05,Ciraci12}. However, for either a conductive spacer\cite{Benz15} or for very small gaps \cite{Savage12,Scholl13} effects due to the increasing conductivity between the two systems have to be considered, with two results: First, the coupled bonding dimer plasmon mode (BDP) blue-shifts due to screening of the local field in the gap, forming the screened coupled mode (SBDP). In this conductivity region the screening field reduces the voltage between the two elements. Second, for higher conductivities a charge transfer mode, characterized by a net current between the two nanoparticles is formed \cite{Perez10}. Since both modes have different physical origins they have to be described by different equivalent circuits. Here we restrict ourselves to the screening of the coupled mode as this is directly accessible in our optical experiments.

\begin{figure}[htbp]
	\centering
	\includegraphics[width=9cm]{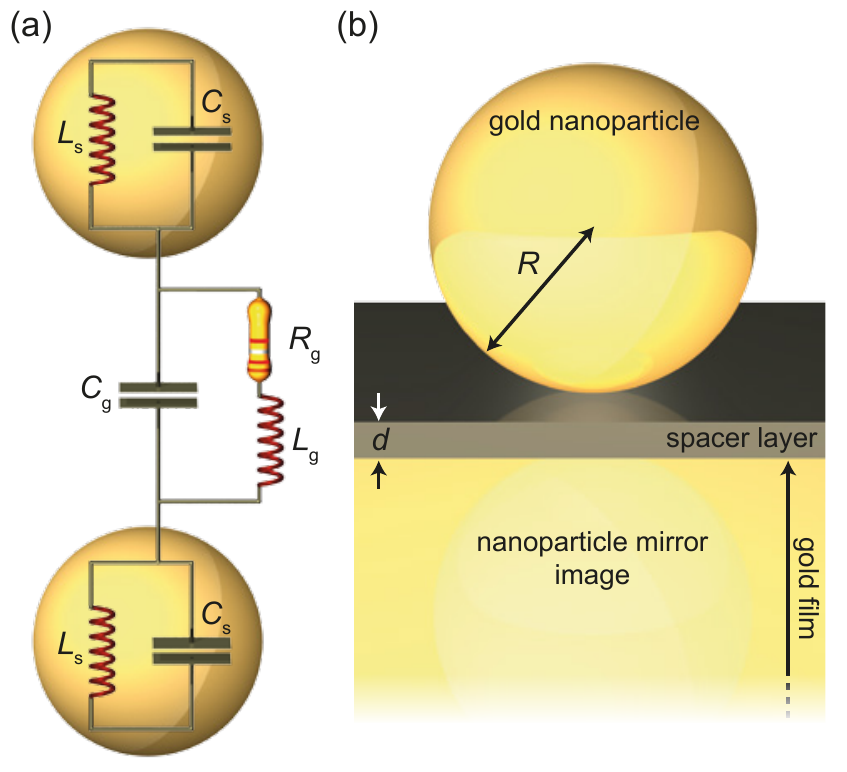}
	\caption{Illustration of the used circuit model and the nanoparticle on mirror geometry. (a) Equivalent circuit for a coupled plasmon mode with both purely capacitive coupling and conductive coupling. (b) Illustration of the nanoparticle on mirror (NPoM) geometry. Source data available in Dataset 1 (Ref.~\cite{OpenData}).}
	\label{Fig:2}
\end{figure}

Previously we considered the ratio of the screened charge to the charge stored on a capacitor without any conductive link \cite{Benz15}. This simplified picture can be described by a resistor in series with the coupling capacitor effectively reducing the voltage at the capacitor. While this approach works well for describing relative changes of the charge stored on the capacitor per optical cycle it is not capable of predicting the absolute value of the resonance wavelength. Therefore we use here an equivalent parallel circuit, which is composed of a capacitor shunted with a resistor and an inductor, which reproduces the behaviour of conductive paths through the gap, in series (see Fig.~\ref{Fig:2}(a)). For a sufficiently low resistance the inductor produces a screening voltage, while for high resistance the inductor plays no role and the coupling is purely dominated by the capacitor. If the gap inductance is large, then little screening is obtained because charge cannot short across the gap within an optical cycle. Hence the fully screened mode at large conductance is blue-shifted furthest if the gap inductance approaches the inductance given by the individual sphere.
The resonance frequency of this circuit found when $\Im (1/Z_{tot})=0$ is

\begin{equation}
\label{Eq:2}
Z_{tot} = \frac{2}{-2 \mathrm{i} \pi \omega R \epsilon_m \epsilon_0 - \mathrm{i} \pi \omega R \epsilon \epsilon_0} + \frac{1}{-\mathrm{i} \omega C_g + \left[R_g-\mathrm{i}\omega L_g \right]^{-1}}.
\end{equation}
\noindent
Simplifying this equation (see appendix~\ref{Ap:A}) yields the resonance condition:

\begin{equation}
\label{Eq:3}
\widetilde{\omega}^4 + \widetilde{\omega}^2 \left[\omega_{d}^{2} \left(\frac{4 \epsilon_m}{\Gamma}-1\right)+\delta\right] - \omega_{d}^{2} \delta = 0
\end{equation}
\noindent
where $\widetilde{\omega} = \omega / \omega_p$ is the normalised resonance frequency, the dimer mode $\omega_d = 1 / \sqrt{2\epsilon_m+\epsilon_\infty+4\epsilon_m \eta}$, normalised gap capacitance $\eta = C_g / C_s$, normalised inductive coupling rate $\Gamma = -L_g C_s \omega_{p}^2$, and normalised loss rate $\delta=\left( \frac{R_g}{L_g \omega_p} \right)^2$. For the case of very low conductivity ($R_g \rightarrow 0$) which gives the typical coupled mode of the bonding dimer plasmon (BDP) this can be simplified to $\widetilde{\omega} = \omega_d$ giving resonant wavelength

\begin{equation}
\label{Eq:4}
\lambda_{BDP} = \lambda_d = \lambda_p \sqrt{2\epsilon_m+\epsilon_\infty+4\epsilon_m \eta}
\end{equation}
\noindent
for coupled plasmonic particles. In the extreme case of high conductivity ($R_g \rightarrow \infty$ so $\delta^{-1} \rightarrow 0$), Eq.~(\ref{Eq:3}) yields the screened coupled plasmon mode (SBDP)

\begin{equation}
\label{Eq:5}
\lambda_{SBDP} = \frac{\lambda_d}{\sqrt{1-\frac{4\epsilon_m}{\Gamma}}} = \lambda_p \sqrt{\frac{2\epsilon_m+\epsilon_\infty+4\epsilon_m \eta}{1-\frac{4\epsilon_m}{\Gamma}}}
\end{equation}
\noindent
which is a central result of our paper.

We now use the nanoparticle on mirror geometry to test this model. In this geometry a gold nanoparticle is placed on a thin spacer layer above a gold film (Fig.~\ref{Fig:2}(b)). Dipoles induced in the nanoparticle can couple to their image charges forming a tightly confined coupled plasmon mode\cite{Nordlander04}. The gap capacitance for this geometry can be obtained following Hudlet \textit{et al.} \cite{Hudlet98} (see appendix~\ref{Ap:B}):

\begin{equation}
\label{Eq:6}
C_g = 2 \pi \epsilon_0 \left(n_g \right)^{\chi} R \times \ln\left[1+ \frac{R}{2d} \theta_{max}^2 \right]
\end{equation}
\noindent
with refractive index of the material in the gap $n_g$, separation between nanoparticle and gold film $d$, and parameters $\theta_{max}$  (which arises from the laterally localised electric field) and $\chi = 1.14$ (which arises from the specific NPoM geometry, appendix~\ref{Ap:B}). To check the validity of the model simulations were performed using the full electrodynamic boundary-element method (BEM) \cite{Abajo97,Abajo02}.

Additional experimental data are obtained by assembling spacers of self-assembled monolayers (SAMs) of aliphatic monothiols $\mathrm{CH}_3 -(\mathrm{CH}_2)_x-\mathrm{SH}$ with $x$ from 3 to 17 on atomically flat template stripped gold \cite{Hegner93}. The thickness of each SAM is measured using spectroscopic ellipsometry. Subsequently, $80\mathrm{nm}$ gold nanoparticles are assembled on top of these SAMs. The resonance wavelengths in air ($\epsilon_m=1$) of more than 1,000 particles per sample were recorded using a fully automated darkfield microscope and cooled spectrometer. Further experimental details can be found in the appendix~\ref{Ap:C} and elsewhere \cite{Nijs15}. The average resonance wavelengths and standard errors were determined by fitting Gaussian distributions to the histograms (see appendix~\ref{Ap:C}). 

\begin{figure}[hp]
	\centering
	\includegraphics[width=12cm]{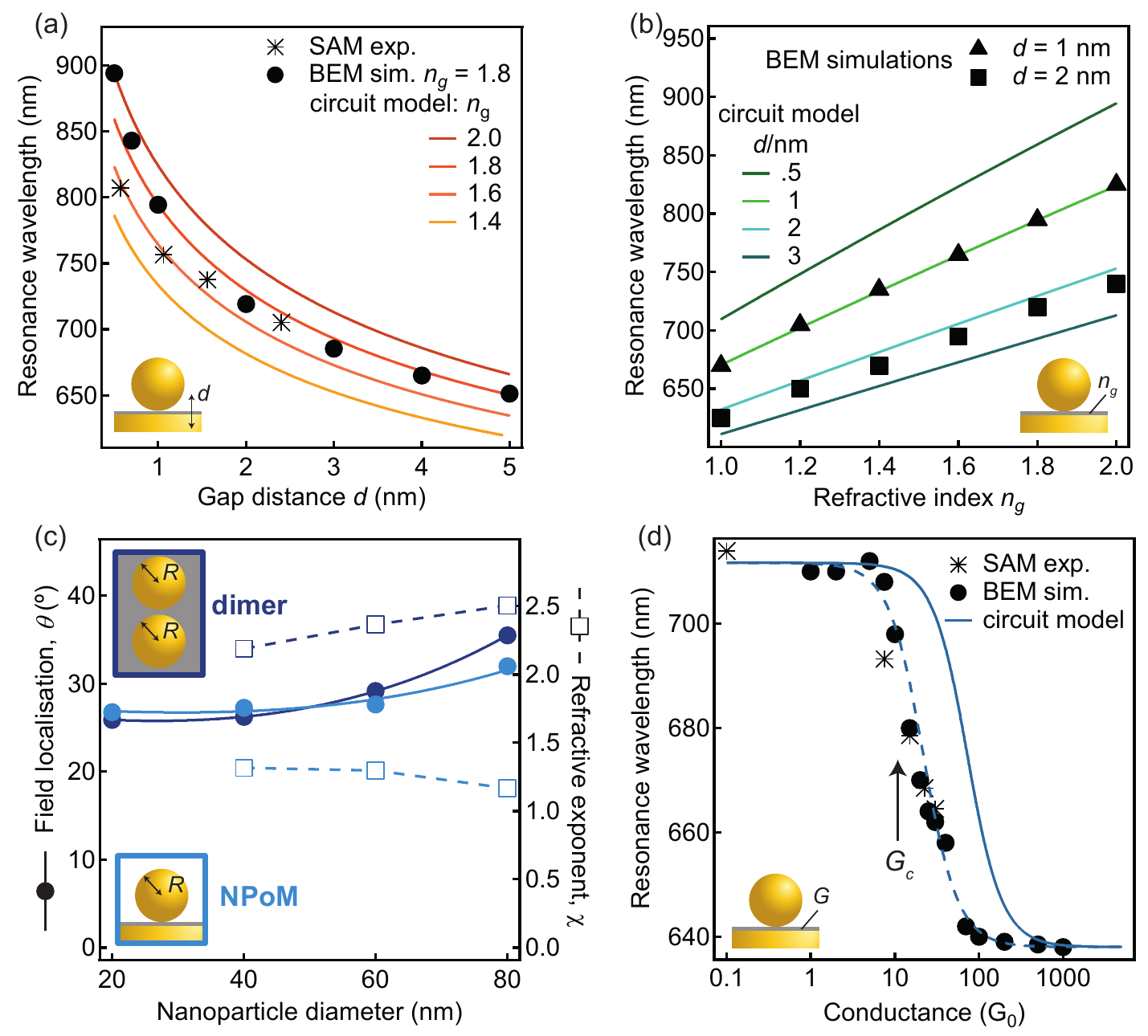}
	\caption{Comparison of the presented circuit model to experimental and simulated data. (a,b) Coupled plasmon mode with $R=40\mathrm{nm}$ NPoM for different gap separations and refractive indices. (c) Nanoparticle size and geometry dependence of the lateral field localisation (solid circles, expressed in terms of the angle $\theta_{max}$, see appendix~\ref{Ap:B}) and refractive index exponent, $\chi$ (empty squares). (d) Blue-shift of the coupled plasmon mode as gap conductance increases, for $R=30\mathrm{nm}$, simulated and experimental data from \cite{Benz15}. The dashed line is shifted by $170 G_0$ (see main text). Source data available in Dataset 1 (Ref.~\cite{OpenData}).}
	\label{Fig:3}
\end{figure}

Comparing our analytical circuit model, BEM simulations, and experimental data shows excellent agreement. The characteristic continuous blueshift of the coupled plasmon resonance for increasing gap spacing and different refractive indices is clearly shown in Fig.~\ref{Fig:3}(a). Similarly the expected linear redshift with increasing refractive index is completely reproduced (Fig.~\ref{Fig:3}(b)).

The gap capacitance parameters are found to be weakly dependent on the nanoparticle size and geometry (Fig.~\ref{Fig:3}(c)). This is because the field localisation (described here by $\theta_{max}$) is slightly different between NPoM and two nanoparticles in a dimer, in particular in the effects of increasing the spacer refractive index (see appendix~\ref{Ap:B}). Most crucially, the model predicts for the first time analytically the blue shift of the coupled plasmon when the gap starts conducting (Fig.~\ref{Fig:3}(d)). While such effects from mixed conducting SAM layers have only previously been computed in full simulations \cite{Perez10}, our analytic solution indeed captures three important properties of such a conductive system: (i) the need to exceed a critical conductance $G_c$, (ii) continuous blue shifts with increasing conductance, and (iii) the saturation of the screening for large conductances. The exact critical conductance is overestimated in our model (Fig.~\ref{Fig:3}(d) solid line), and fits much better when shifted by $-170 G_0$ (dashed Fig.~\ref{Fig:3}(d)), however the order of magnitude is acceptable. Using Eq.~(\ref{Eq:3}), we obtain an analytical estimate for the critical conductance, $\frac{G_c}{G_0} \simeq \frac{4\pi^2}{G_0 Z_0} \frac{R}{\lambda_p} \frac{\lambda_d}{\Gamma}$  where $Z_0 = 377\Omega$ is the impedance of free space (see appendix~\ref{Ap:A}).

\section{The role of the gap inductance}

\begin{figure}[htbp]
	
	\centering
	\includegraphics[width=12cm]{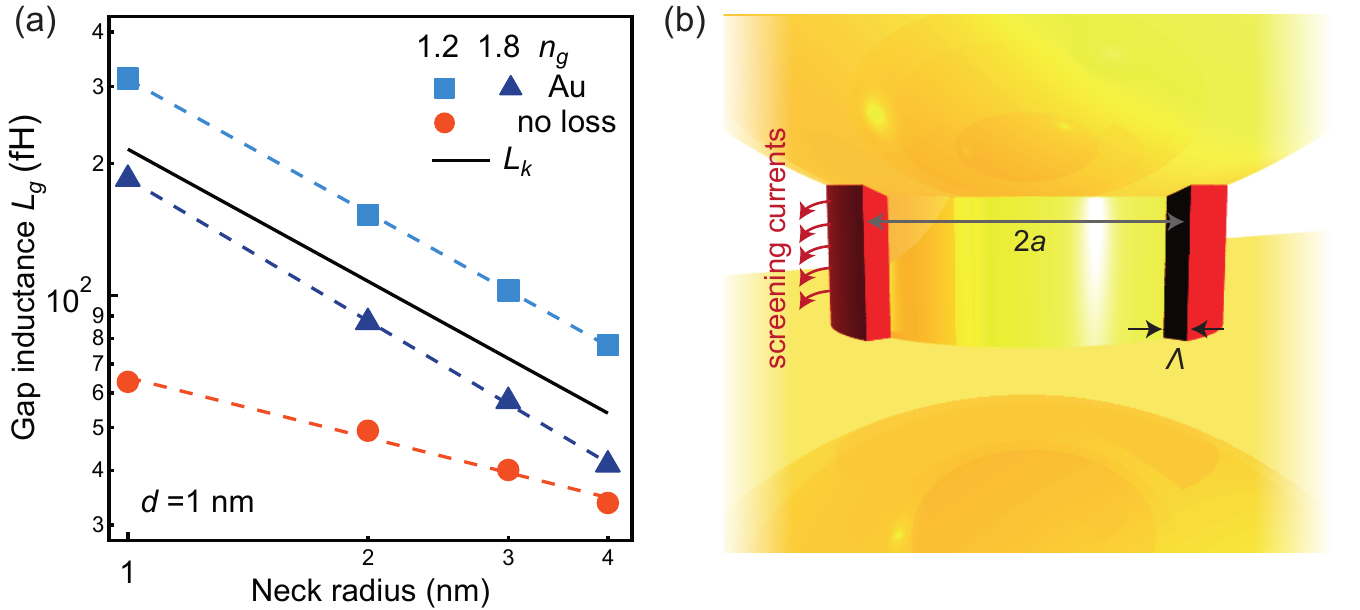}
	\caption{Interpretation of the introduced gap inductance. (a) Gap inductance vs neck radius, extracted by using Eq.~(\ref{Eq:5}) on simulation results, and compared to kinetic inductance model (solid line). Filaments are either gold, or near-perfect conductor. (b) Schematic of sheath currents confined to penetration depth $\Lambda$ down conducting filament, which gives kinetic inductance. Source data available in Dataset 1 (Ref.~\cite{OpenData}).}
	\label{Fig:4}
\end{figure}

The maximum blue shift when the gap becomes conducting (saturating the screening) is set by the inductance of this filament, which screens charges in the gap expelling the electric field. To understand what sets the value of this inductance we perform BEM simulations for both insulating and completely conducting gaps. The conductive region was modelled as a cylinder connecting the gold spheres. Varying the neck radius a reveals that $L_g \propto a^{-1}$ for gold while for an ideal conductor (with extremely low damping) we find $L_g \propto a^{-\frac{1}{2}}$ (Fig.~\ref{Fig:4}(a)). This dependence arises because $L_g$ is dominated by the kinetic inductance $L_k$ produced by the magnetic field in the gap encircling the current acting to drag back the electrons. To obtain $L_k$ we equate the kinetic energy of $N$ moving electrons with the energy of an inductor $N \frac{1}{2} m v^2 = \frac{1}{2} L_k I^2$ with velocity $v$, current $I=nevA$, area $A$, and $N=nAd$, giving $L_k= \frac{d}{A} \frac{m}{ne^2} = \frac{d}{A \epsilon_0 \omega_p^2}$ with density of mobile carriers in the conducting link $n$ and electron mass $m$ \cite{Staffaroni12}. The electric field however can only penetrate a certain distance $\Lambda$ into the connecting filament so that current flows only in an outer sheath (Fig.~\ref{Fig:4}(b)) giving

\begin{equation}
\label{Eq:7}
L_k = \frac{1}{\epsilon_0 \omega_p^2} \frac{d+2\Lambda}{2\pi \Lambda a} = \frac{1}{\epsilon_0 \omega_p^2} \frac{f}{a}
\end{equation}

\noindent
with the geometry factor $f= \frac{1}{\pi} \left[ 1+\frac{d}{2 \Lambda}\right]$. We find screening depths of $\Lambda \sim 3\mathrm{nm}$ (appendix~\ref{Ap:A}) so for very narrow gold gaps, $f \rightarrow \frac{1}{\pi}$  which then accounts extremely well for the $a^{-1}$ dependence in the full simulations of $L_g$ and even produces quantitative agreement with the magnitude of the gap inductance (Fig.~\ref{Fig:4}(a)). For an imperfect metal the screening depth cannot be ignored, giving a modified form which depends also on $n_g$ (appendix~\ref{Ap:A}). The blue-shift of the coupled plasmon seen when conducting links are inserted between plasmonic elements can thus be understood as arising from their kinetic inductance. Inserting Eq.~(\ref{Eq:7}) into Eq.~(\ref{Eq:5}), we find an analytic form of the blue-shift

\begin{equation}
\label{Eq:8}
\frac{\Delta \lambda}{\lambda} = \epsilon_m \frac{a}{R} \left( 1 + \frac{d}{2 \Lambda}\right)^{-1}
\end{equation}

\noindent
which for small gaps depends linearly on the ratio of neck to nanoparticle radius. These shifts can indeed be large (we obtain $70\mathrm{nm}$ spectral shifts experimentally, see Fig.~\ref{Fig:3}(d)). We also note the interesting ramification that $\Delta \lambda$ will depend on the morphology of links across a plasmonic gap. This opens great opportunities in understanding nanoscale conducting elements purely through optical spectroscopy, which can then follow dynamic processes.

\section{Conclusion}
In conclusion, we present a circuit model which can be used to describe coupled plasmon resonances in both capacitive and conductive coupling regimes. It provides an exceptionally straightforward way to calculate analytically the resonance wavelength for different gap sizes, nanoparticle sizes, refractive indices, and linker conductivities. We showed how a test geometry based on a nanoparticle on gold mirror gives excellent agreement of the model with both experiment and simulation. This model allows us to identify the crucial properties of the system, and resolves the key role of kinetic inductance. Further geometries can easily be incorporated by adjusting the capacitances of both the gap region and the individual plasmonic systems.

\section*{Appendix}
\appendix

\section{Derivation of the model}\label{Ap:A}
The resonance frequency of the circuit shown in Fig.~\ref{Fig:2}(a) (main text) is found when $\Im (1/Z_{tot})=0$:

\begin{equation}
\label{Eq:S1}
Z_{tot} = \frac{2}{-2 \mathrm{i} \pi \omega R \epsilon_m \epsilon_0 - \mathrm{i} \pi \omega R \epsilon \epsilon_0} + \frac{1}{-\mathrm{i} \omega C_g + \left[R_g-\mathrm{i}\omega L_g \right]^{-1}}
\end{equation}

\begin{equation}
\label{Eq:S2}
Z_{tot} = \frac{4\mathrm{i}}{2\pi R \epsilon_{0}} \times \frac{\omega}{\omega^2 \left( 2\epsilon_m + \epsilon \right)} + \frac{\mathrm{i}}{C_s} \times \frac{1}{\omega \frac{C_g}{C_s} + \frac{1}{-\omega C_s L_g - \mathrm{i} R_g C_s}}
\end{equation}
\noindent
Using the Drude model Eq.~(\ref{Eq:1}) for $\epsilon$ and replacing all frequencies $\omega$ by $\widetilde{\omega} = \omega / \omega_p$ yields:

\begin{equation}
\label{Eq:S3}
Z_{tot} = \frac{\mathrm{i}}{2 \pi R \omega_p \epsilon_0} \left( \frac{4 \omega}{\left(2 \epsilon_m + \epsilon_{\infty}\right) \widetilde{\omega}^2 -1} + \frac{1}{\epsilon_m} \frac{1}{\widetilde{\omega}\eta + \frac{1}{\widetilde{\omega}\Gamma - \mathrm{i} \gamma}} \right) 
\end{equation}
\noindent
with $\eta = \frac{C_g}{C_s}$ , $\Gamma= -L_g C_s \omega_p^2$, and $\gamma = R_g C_s \omega_p$. Rearranging Eq.~(\ref{Eq:S3}) yields:

\begin{equation}
\label{Eq:S4}
Z_{tot} = \frac{1}{- \mathrm{i} 2 \pi R \omega_p \epsilon_0} \left( \frac{4 \widetilde{\omega} \epsilon_m \left( \widetilde{\omega} \eta + \frac{1}{\widetilde{\omega}\Gamma - \mathrm{i} \gamma}\right) + \left( 2 \epsilon_m + \epsilon_{\infty} \right) \widetilde{\omega}^2 -1}
{\epsilon_m \left(\widetilde{\omega} \eta + \frac{1}{\widetilde{\omega} \Gamma - \mathrm{i} \gamma}\right)\left[ \left( 2 \epsilon_m + \epsilon_{\infty}\right)\widetilde{\omega}^2 -1\right]}  \right) 
\end{equation}

\begin{equation}
\label{Eq:S5}
\Im \left( \frac{1}{Z_{tot}} \right) = 
\Im \left[
-\mathrm{i} 2 \pi R \omega_p \epsilon_0
\left(
\frac{\epsilon_m \left(\widetilde{\omega} \eta + \frac{1}{\widetilde{\omega} \Gamma - \mathrm{i} \gamma} \right) \left[ \left(2 \epsilon_m + \epsilon_{\infty} \right) \widetilde{\omega}^2 -1\right]}
{4 \widetilde{\omega} \epsilon_m \left(\widetilde{\omega} \eta + \frac{1}{\widetilde{\omega} \Gamma - \mathrm{i} \gamma} \right)
	+ \left(2 \epsilon_m + \epsilon_{\infty} \right)\widetilde{\omega}^2 -1}
\right)
\right]
\end{equation}

\begin{equation}
\label{Eq:S6}
\Im \left( \frac{1}{Z_{tot}} \right) = 
\Re \left[
- 2 \pi R \omega_p \epsilon_0
\left(
\frac{\epsilon_m \left(\widetilde{\omega} \eta + \frac{1}{\widetilde{\omega} \Gamma - \mathrm{i} \gamma} \right) \left[ \left(2 \epsilon_m + \epsilon_{\infty} \right) \widetilde{\omega}^2 -1\right]}
{4 \widetilde{\omega} \epsilon_m \left(\widetilde{\omega} \eta + \frac{1}{\widetilde{\omega} \Gamma - \mathrm{i} \gamma} \right)
	+ \left(2 \epsilon_m + \epsilon_{\infty} \right)\widetilde{\omega}^2 -1}
\right)
\right] \overset{!}{=} 0
\end{equation}
\noindent
To look for the real part of this expression we multiply with the complex conjugate of the denominator, and set the numerator to zero:

\begin{align}
\begin{aligned}
\label{Eq:S7}
\Re \biggl[
\epsilon_m \left(\widetilde{\omega} \eta + \frac{\widetilde{\omega} \Gamma + \mathrm{i} \gamma}{\widetilde{\omega}^2 \Gamma^2 + \gamma^2} \right)
\left[\left(2\epsilon_m + \epsilon_{\infty}\right)\widetilde{\omega}^2 -1\right] \\
\left[4 \widetilde{\omega} \epsilon_m \left(\widetilde{\omega} \eta + \frac{\widetilde{\omega}\Gamma - \mathrm{i} \gamma}{\widetilde{\omega}^2 \Gamma^2 + \gamma^2}\right)
+ \left(2\epsilon_m + \epsilon_{\infty}\right)\widetilde{\omega}^2 - 1
\right]
\biggr] \overset{!}{=} 0
\end{aligned}
\end{align}

\begin{equation}
\label{Eq:S8}
\Re \left[
\epsilon_m \left(
\widetilde{\omega} \eta +
\frac{\widetilde{\omega}\Gamma + \mathrm{i} \gamma}{\widetilde{\omega}^2 \Gamma^2 + \gamma^2}\right)
\left[4 \widetilde{\omega} \epsilon_m \left(
\widetilde{\omega} \eta +
\frac{\widetilde{\omega}\Gamma - \mathrm{i} \gamma}{\widetilde{\omega}^2 \Gamma^2 + \gamma^2}
\right)
+ \left(2 \epsilon_m + \epsilon_{\infty} \right) \widetilde{\omega}^2 - 1
\right]
\right] \overset{!}{=} 0
\end{equation}

\begin{align}
\begin{aligned}
\label{Eq:S9}
\epsilon_m \left(
\widetilde{\omega} \eta +
\frac{\widetilde{\omega}\Gamma}{\widetilde{\omega}^2 \Gamma^2 + \gamma^2}\right)
\left[
\left(2 \epsilon_m + \epsilon_{\infty}\right) \widetilde{\omega}^2 - 1 + 4 \widetilde{\omega} \epsilon_m \left( \widetilde{\omega} \eta +
\frac{\widetilde{\omega}\Gamma}{\widetilde{\omega}^2 \Gamma^2 + \gamma^2}\right)
\right] \\
 + \left(\frac{\gamma \epsilon_m}{\widetilde{\omega}^2 \Gamma^2 + \gamma^2}\right)
\left(\frac{4 \widetilde{\omega} \gamma \epsilon_m}{\widetilde{\omega}^2 \Gamma^2 + \gamma^2}\right)
\overset{!}{=} 0
\end{aligned}
\end{align}

\begin{align}
\begin{aligned}
\label{Eq:S10}
\epsilon_m \widetilde{\omega} \left(\eta \left( \widetilde{\omega}^2 \Gamma^2 + \gamma^2 \right) + \frac{\Gamma}{1}\right)
\left[
\left(2\epsilon_m + \epsilon_{\infty} \right) \widetilde{\omega}^2
- 1 + 4 \widetilde{\omega} \epsilon_m \left(\widetilde{\omega} \eta + \frac{\widetilde{\omega} \Gamma}{\widetilde{\omega}^2 \Gamma^2 + \gamma^2}\right)
\right] + \\
\gamma \epsilon_m \left( \frac{4\widetilde{\omega}\gamma\epsilon_m}{\widetilde{\omega}^2 \Gamma^2 + \gamma^2} \right) \overset{!}{=} 0
\end{aligned}
\end{align}

\begin{align}
\begin{aligned}
\label{Eq:S11}
\widetilde{\omega}^2 \epsilon_m \left[\eta \left(\widetilde{\omega}^2 \Gamma^2 + \gamma^2 \right) + \Gamma \right]^2 + \gamma^2 \epsilon_m + \\
\frac{\left(2 \epsilon_m + \epsilon_{\infty} \right)\widetilde{\omega}^2 - 1}{4}
\left[\eta \left(\widetilde{\omega}^2 \Gamma^2 + \gamma^2 \right) + \Gamma \right]\left( \widetilde{\omega}^2 \Gamma^2 + \gamma^2 \right)
\overset{!}{=} 0
\end{aligned}
\end{align}
\noindent
We find to good approximation that $\gamma^2 << \widetilde{\omega}^2 \Gamma^2$, simplifying Eq.~(\ref{Eq:S11}) further. Using $\omega_d = 1 / \sqrt{2\epsilon_m + \epsilon_{\infty} + 4 \epsilon_m \eta} = \lambda_d^{-1}$ and $\delta = \left(\frac{\gamma}{\Gamma}\right)^2$ yields:

\begin{equation}
\label{Eq.S12}
\widetilde{\omega}^4 + \widetilde{\omega}^2 \left[\omega_{d}^{2} \left(\frac{4 \epsilon_m}{\Gamma}-1\right)+\delta\right] - \omega_{d}^{2} \delta = 0
\end{equation}
\noindent
This is Eq.~(\ref{Eq:3}) in the main text. The solution to this gives the spectral position of the dimer resonance, given suitable parameters.

\subsection{Conductive gaps}
\begin{figure}[htbp]
	\centering
	\includegraphics[width=6cm]{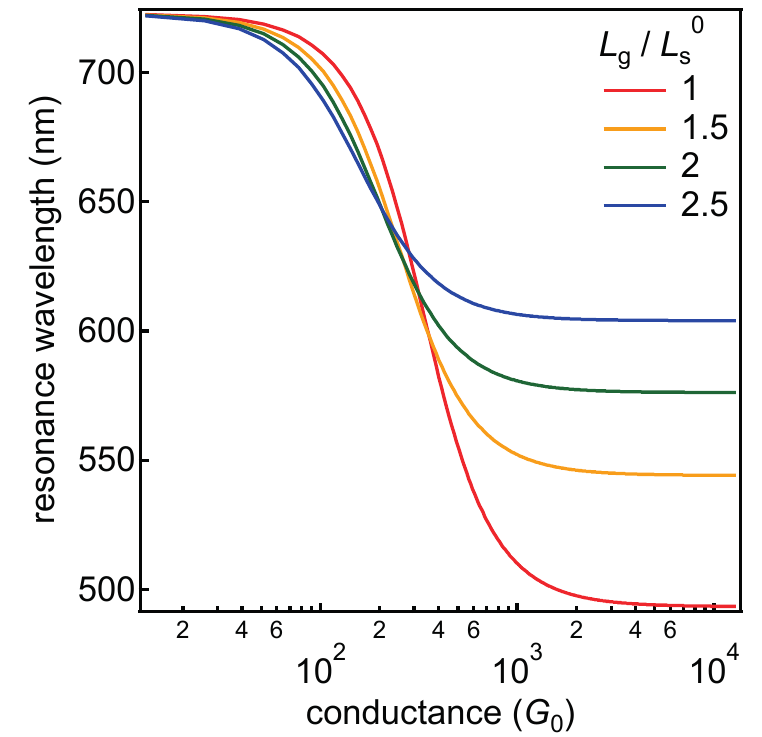}
	\caption{Resonance wavelength as a function of the gap conductivity for several gap inductances (80 nm NPoM, $d$ = 1.1 nm, $n$ = 1.45). Source data available in Dataset 1 (Ref.~\cite{OpenData}).}
	\label{Fig:S1}
\end{figure}

In the case of conductive gaps, the resistance $R_g$ controls the crossover between the coupled resonance and the screened coupled resonance. The inductance $L_g$ determines the spectral position of the fully screened mode (\textit{i.e.} highly conducting gap where $\delta$ becomes small), so that $\frac{\lambda_{SBDP}}{\lambda_p}=\lambda_d \left(1-\frac{4\epsilon_m}{\Gamma}\right)^{-1/2}$. We write this in terms of the ratio between gap and sphere inductance, $\Gamma=-\frac{L_g}{L_s^0} \left(2+\epsilon_{\infty}\right)$, with $L_s^0$ taken at the single particle resonance. If the gap inductance is large, then little screening is obtained because charge cannot short across the gap within an optical cycle. Hence the fully screened mode at large conductance is blue-shifted furthest if the gap inductance approaches the inductance given by the individual sphere ($\frac{L_g}{L_s^0} = 1$). The cross-over conductance between the two regimes is given by $\gamma=-\Gamma\omega_d \sqrt{1-\frac{2}{\Gamma}}$ which implies a critical conductance $G_c=\omega_p C_s \frac{\lambda_d}{\sqrt{\Gamma^2+2\Gamma}}=1360 \frac{R}{\lambda_p} \frac{\lambda_d}{\sqrt{\Gamma^2+2\Gamma}} G_0$. This predicts indeed the right magnitude of the critical conductance compared to the full BEM simulations. Changing the gap inductance thus modifies both the wavelength of the screened mode, as well as the critical conductance (Fig.~\ref{Fig:S1}).

For the kinetic inductance as described in the main text, we can substitute into
\begin{align}
\begin{aligned}
\label{Eq:S13}
\lambda_{SBDP} = \frac{\lambda_d}{\sqrt{1-\frac{4\epsilon_m}{\Gamma}}} \\
= \frac{\lambda_d}{\sqrt{1-\frac{4\epsilon_m}{C_s L_g \omega_p^2}}}=\frac{\lambda_d}{\sqrt{1-\frac{4\epsilon_m\epsilon_0\omega_p^2 a}{2\pi\epsilon_0 R f\omega_p^2}}}=\frac{\lambda_d}{\sqrt{1-\frac{2\epsilon_m a}{\pi R f}}}
\end{aligned}
\end{align}
\noindent
using $\Gamma=L_g C_s \omega_p^2$ with $C_s=2\pi\epsilon_0 R$, $L_g=\frac{1}{\epsilon_0\omega_p^2} \frac{f}{a}$, and then using a binomial expansion we find:

\begin{equation}
\label{Eq:S14}
\lambda_{SBDP}=\lambda_d \left(1+\frac{\epsilon_m a}{\pi R f}\right)
\end{equation}
\noindent
which gives the result for the blue shift in the main text.

\subsection{Screening depths}
In the small-gap metal-insulator-metal structures here, we find the propagation wavevector in a gap of width $d$

\begin{equation}
\label{Eq:S15}
S_m=\frac{2\epsilon_d}{d \epsilon_m}
\end{equation}

which gives the field in the metal 
\begin{equation}
\label{Eq:S16}
E_z \propto \exp{\left(S_m z\right)}
\end{equation}
Solving for the penetration depth of optical field inside the metal
\begin{equation}
\label{Eq:S17}
\Lambda=\frac{1}{\Re \left(S_m\right)}=\frac{d}{2\epsilon_d \Re {\left(1/\epsilon_m\right)}}
\end{equation}

\begin{figure}[htbp]
	\centering
	\includegraphics[width=6cm]{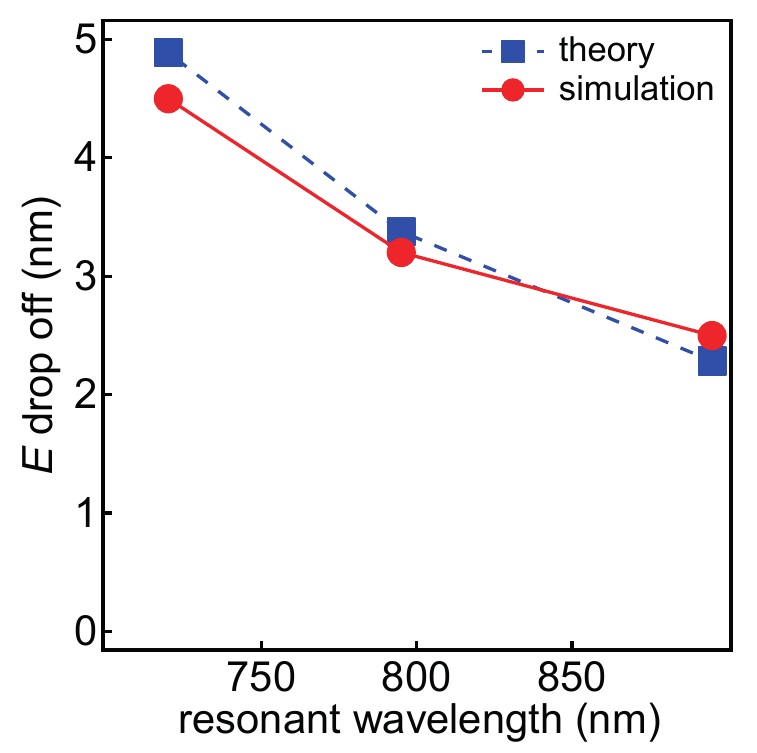}
	\caption{Field penetration depth in NPoM. Exponential decay length of optical field calculated from simple model above, and from full BEM simulations. Source data available in Dataset 1 (Ref.~\cite{OpenData}).}
	\label{Fig:S2}
\end{figure}

We compare below this penetration depth (Fig.~\ref{Fig:S2} blue squares) with full BEM simulations of the exact NPoM structures (spherical 80nm Au NP) from which we also extract the field penetration (Fig.~\ref{Fig:S2} red circles). This gives a very good match for the various ($d$ [nm],$n_g$) = (1,1.3), (1,1.8), (0.5,1.8) which shift the resonant mode wavelength across the near infrared. 

We can also substitute in the Drude model, and providing the damping is not too large we find

\begin{equation}
\label{Eq:S18}
\Lambda=\frac{d\lvert\epsilon_{\infty}-\frac{\lambda^2}{\lambda_p^2}\rvert}{2 n_g^2}
\end{equation}

\section{Nanoparticle on mirror capacitance model}\label{Ap:B}
\begin{figure}[htbp]
	\centering
	\includegraphics[width=12cm]{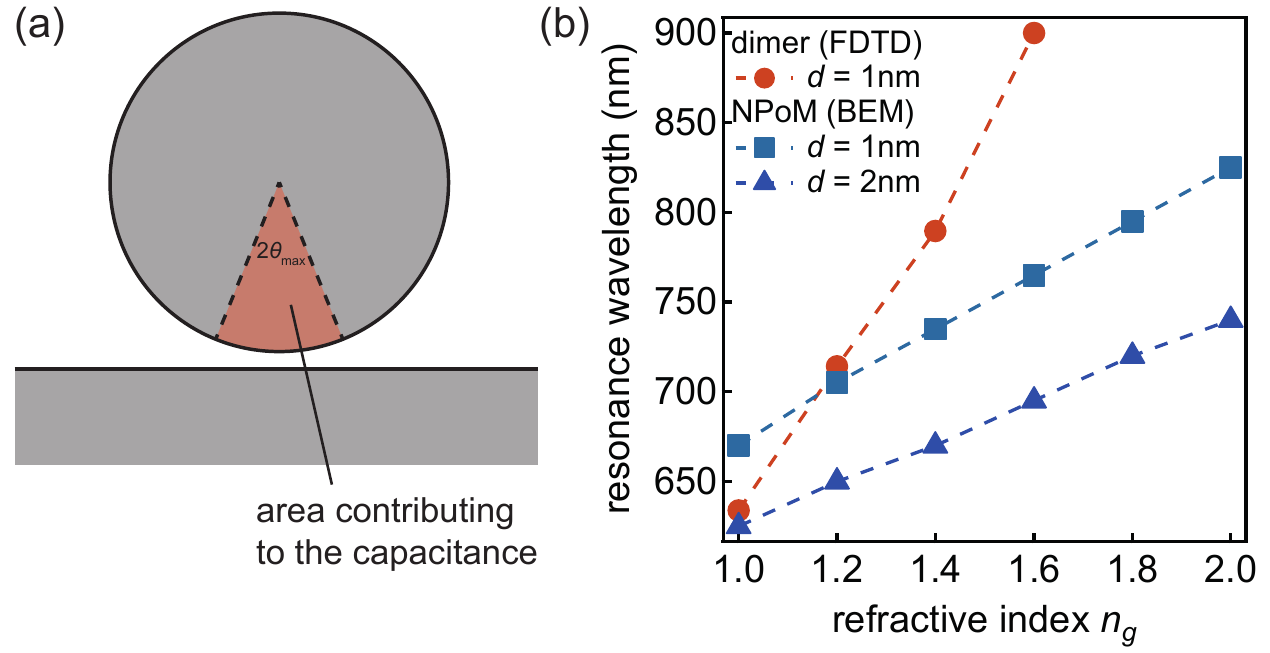}
	\caption{Illustration of the used capacitance model. (a) By using a $\theta_{max}$  much smaller than $\pi$, only a part of the sphere is taken into account for the capacitance. Throughout the paper $\theta_{max}\simeq\pi / 5$ is used for the nanoparticle on mirror geometry. (b) Refractive index dependency for both nanoparticle on mirror and nanoparticle dimers, illustrating the different dependencies on the refractive index. Source data available in Dataset 1 (Ref.~\cite{OpenData}).}
	\label{Fig:S3}
\end{figure}

Modelling the gap capacitance is key to obtain reliable values for the resonance frequency and for obtaining the correct dependencies on distance and refractive index. The simplest approximation is a plate capacitor, with area $A$, filled with a material with refractive index $n_g$:

\begin{equation}
\label{Eq:S19}
\eta=\frac{C_g}{C_s}=\frac{\epsilon_0 n_g^2 A/d}{2\pi R\epsilon_0} =n_g^2 \frac{A}{2\pi Rd}
\end{equation}
\noindent
A more precise description of the capacitance of a spherical nanoparticle on a metallic mirror is\cite{Hudlet98}:
\begin{equation}
\label{Eq:S20}
\eta= \frac{2 \pi \epsilon_0 n_g^2 R}{2\pi\epsilon_0 R} \int_{0}^{\pi} {\frac{\sin^2{\theta}}{\theta[\frac{d}{R}+1-\cos{\theta}]}d\theta}
\end{equation}
\noindent
By considering that the gap occupies only a small area of the full sphere (\textit{i.e.} restricting the upper integration boundary to a small angle $\theta_{max}$) and using small angle approximations yields (see illustration in Fig.~\ref{Fig:S3}(a)):

\begin{equation}
\label{Eq:S21}
\eta =n_g^2 \ln{\left(1+\frac{R}{2d}\theta_{max}^2\right)}
\end{equation}

\begin{figure}[htbp]
	\centering
	\includegraphics[width=12cm]{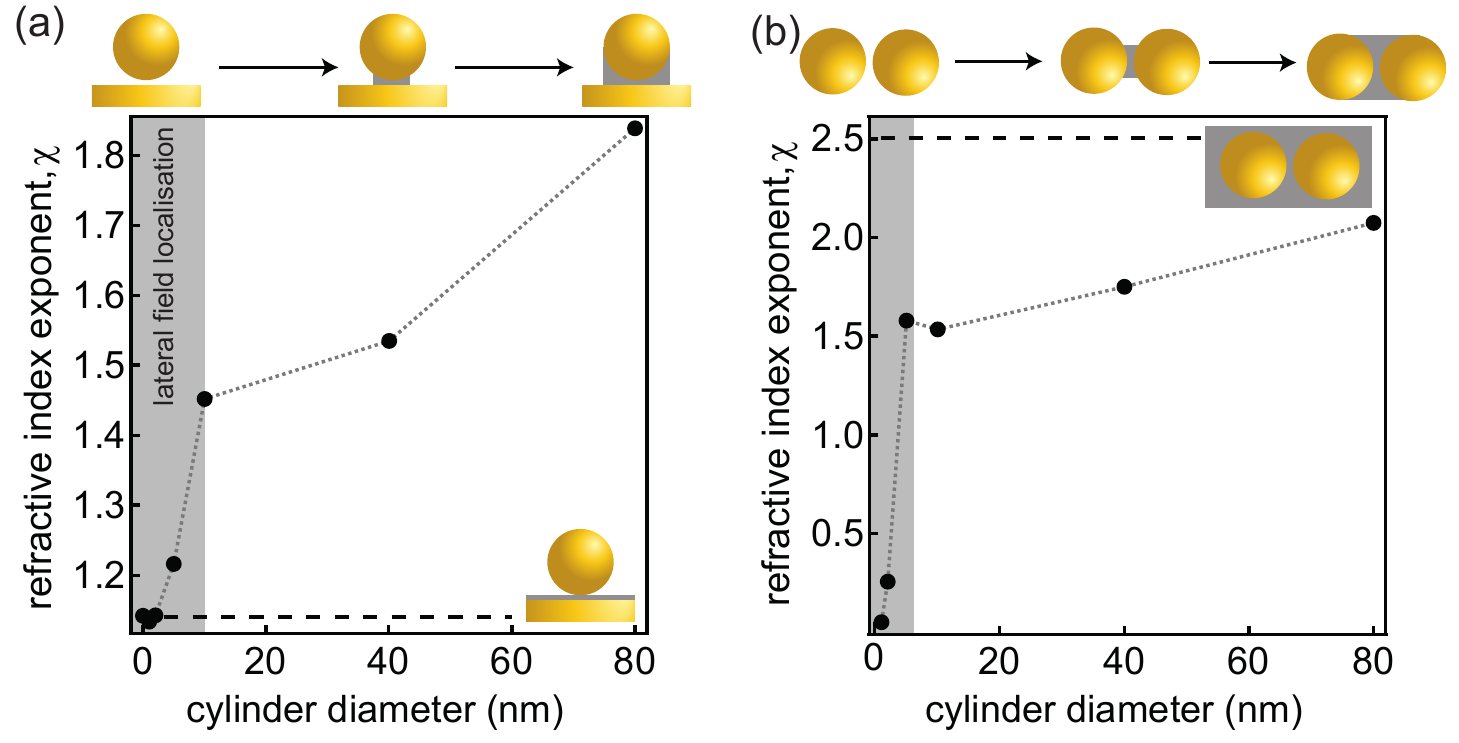}
	\caption{Refractive index exponent $\chi$ for different geometries. (a,b) Nanoparticle on mirror (NPoM) and dimer geometry. The grey shaded area shows the approximate lateral dimension of the plasmon. The dashed lines indicate the exponent value for our two reference geometries: a film covering the whole gold substrate for the nanoparticle on mirror geometry and a nanoparticle dimer submerged in a medium with the respective refractive index for the dimers. Source data available in Dataset 1 (Ref.~\cite{OpenData}).}
	\label{Fig:S4}
\end{figure}

By comparing the refractive index dependency obtained from electromagnetic simulations (finite-difference time-domain - FDTD \& boundary element method - BEM) we find that depending on the exact geometry the exponent of the refractive index deviates from two. We have performed simulations for different geometries and different refractive indices. Plotting $\log{⁡\left(\eta\right)}$ (calculated from the resonance wavelength using Eq.~(\ref{Eq:4})) vs $\log{⁡\left(n\right)}$ allows to extract the refractive index exponent $\chi$. Fig.~\ref{Fig:S4} shows $\chi$ for both nanoparticle dimers and the nanoparticle on mirror geometry.

Fig.~\ref{Fig:S4} shows that using a cylindrical material in the gap can strongly influence the refractive index dependency; we observe two regions: For cylinder diameters smaller than the lateral field localization a steep increase is observed indicating that both the single particle mode and the coupled mode shift their spectral position. A second region is observed for diameter larger than the plasmon dimension, here the dependence is shallow as mostly the single particle mode is changed. For an extended film covering the whole gold substrate we find $\chi =1.14$, for dimers submerged in a medium with a certain refractive $\chi = 2.5$.

\section{Experimental details \& results}\label{Ap:C}
\begin{figure}[htbp]
	\centering
	\includegraphics[width=12cm]{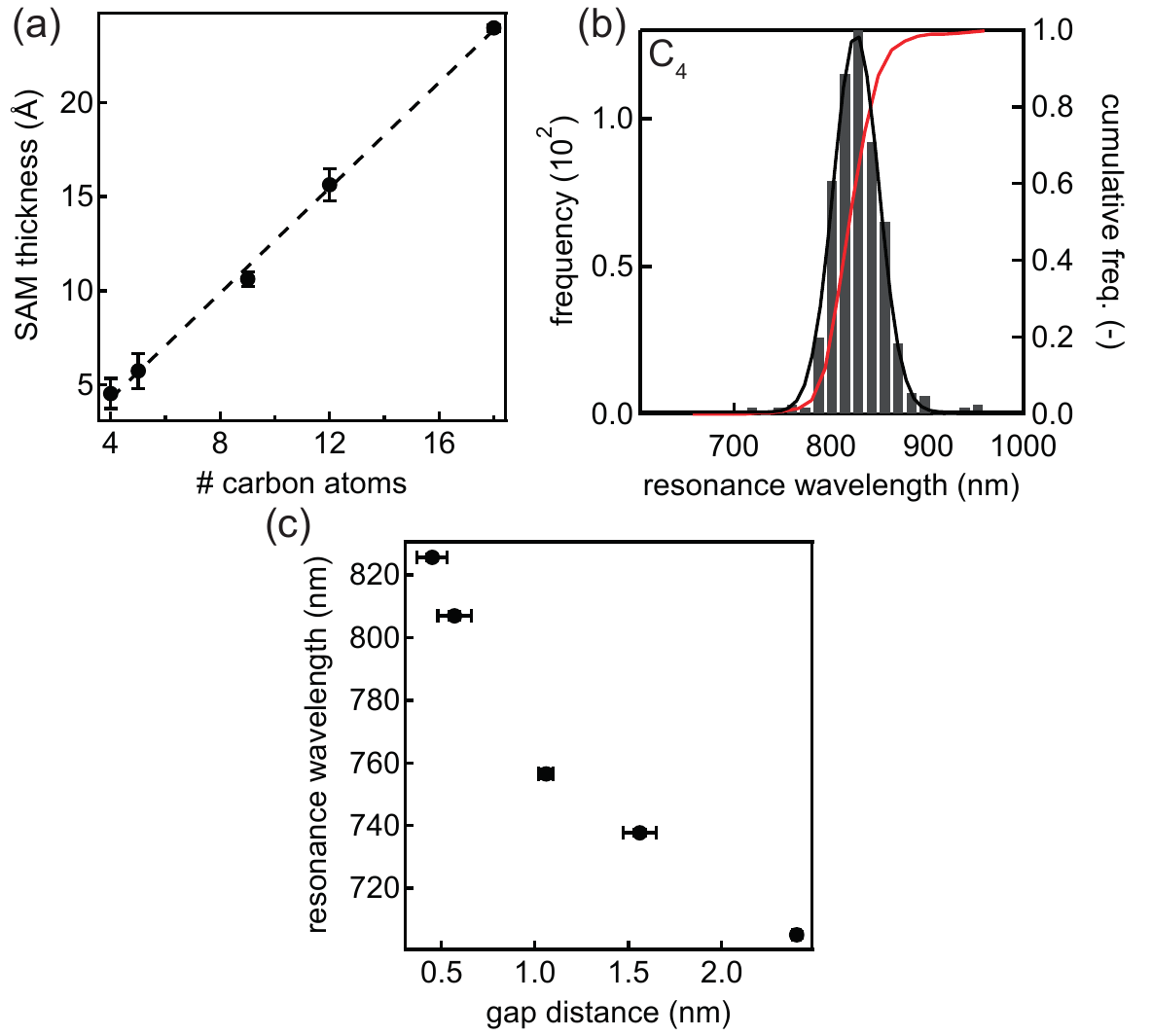}
	\caption{Darkfield spectroscopy of aliphatic SAM samples. (a) SAM thickness as a function of the number of carbon atoms, measured by ellipsometry. (b) Typical distribution of plasmon resonance wavelengths for over 1,000 nanoparticles on a mirror with butanethiol (C$_4$) as a spacer. (c) Resonance wavelength for different aliphatic SAM spacer. Source data available in Dataset 1 (Ref.~\cite{OpenData}).}
	\label{Fig:S5}
\end{figure}

As described in the main text self-assembled monolayers of different length aliphatic thiols were used as experimental spacer. For each SAM the thickness of three samples (three spots per sample) were measured by phase modulated ellipsometry. Fig.~\ref{Fig:S5}(a) shows the determined SAM thickness as a function of the number of carbon atoms. For each SAM a sample was covered with 80 nm gold nanoparticles. The resonance wavelength of over 1,000 nanoparticles was measured per sample using an automated Olympus BX51 darkfield microscope. Further details on the spectroscopy and evaluation can be found elsewhere \cite{Nijs15}. Fig.~\ref{Fig:S5}(b) shows exemplarily an obtained resonance wavelength distribution for butanethiol (C$_4$) self-assembled monolayer as a spacer layer. The resonance wavelengths for all samples as a function of the SAM thickness can be found in Fig.~\ref{Fig:S5}(c).

\subsection{Sample Preparation}
A 100 nm thick gold film was evaporated on a silicon (100) wafer (Kurt J. Lesker Company, PVD 200) using an evaporation rate of 1 \AA/s with a base pressure of $\approx 1\times 10^{-7}$ mbar. To obtain atomically smooth gold surfaces a standard template stripping method was applied: small silicon (100) pieces were glued to the evaporated gold surface using Epo-Tek 377 epoxy glue. After curing the glue the sample can be pealed of the silicon wafer revealing a fresh, clean surface with excellent surface roughness \cite{Hegner93}. 1-butanethiol (C$_4$), 1-pentanethiol (C$_5$), 1-nonanethiol (C$_9$), 1-dodecanethiol (C$_{12}$), and 1-octadecanethiol (C$_{18}$) (Sigma-Aldrich, $\geq98\%$) self-assembled monolayers were assembled on these freshly prepared surfaces by immersing in 1 mM solution for $\approx$22 h. Subsequently, the samples were thoroughly rinsed with ethanol, briefly cleaned in an ultrasonic bath to remove excess unbound thiols, and blown dry. The samples were stored under a constant nitrogen flow until they were used. 80 nm gold nanoparticles were used as received from BBI Solutions. To deposit the nanoparticles on the SAM the previously prepared samples were immersed in the nanoparticle solution. The time was adjusted in order to reach a uniform but sparse coverage that allows spectroscopic investigations of individual nanoparticles. Excess nanoparticles were rinsed off with distilled water and the samples were dried with nitrogen.

\subsection{Experimental}
Darkfield spectra of over 1,000 nanoparticles were recorded using a fully automated Olympus BX51 microscope in a reflective dark field geometry. The scattered light was collected with a $\times$100 long working distance objective (NA 0.8) and analysed with a TEC-cooled Ocean Optics QE65000 spectrometer. Further details can be found elsewhere \cite{Nijs15}.

\subsection{Ellipsometry}
The thickness of the self-assembled monolayers was measured using a Jobin-Yvon UVISEL spectroscopic ellipsometer at angle of incidence 70$^\circ$ and wavelengths from 300-800 nm. The base layer permittivity is obtained from a clean gold substrate and the dielectric properties of the SAMs described using a Cauchy model with refractive index of 1.45 (as determined by \cite{Folkers92,Bain89}).

\section*{Acknowledgments}
We acknowledge financial support from EPSRC grant EP/G060649/1, EP/I012060/1, EP/L027151/1, EP/K028510/1, ERC grant LINASS 320503. FB acknowledges support from the Winton Programme for the Physics of Sustainability. RC wishes to acknowledge financial support from St. John's College, Cambridge for Dr. Manmohan Singh Scholarship. CT and JA acknowledge financial support from Project FIS2013- 41184-P from MINECO, ETORTEK 2014-15 of the Basque Department of Industry and IT756-13 from the Basque consolidated groups.
\end{document}